\documentclass[10pt,english,aps,prl,showpacs,twocolumn]{revtex4}
\usepackage{graphicx}
\usepackage{epsfig}

\begin{document}

\title{Pairing and superconductivity driven by
 strong quasiparticle renormalization in
two-dimensional organic charge transfer salts}
\author{ Jun Liu$^{1}$, J\"{o}rg Schmalian$^{1}$, and Nandini Trivedi$^{2,3}$} \affiliation{$^{1}$Department of Physics and Astronomy and Ames Laboratory, Iowa State University, Ames, Iowa 50011, USA\\ $^{2}$Department of
Theoretical Physics, 
Tata Institute of Fundamental Research, Mumbai 400005, 
India \\ $^{3}$ Department of Physics, Ohio State University,  
Columbus, Ohio 43210, USA} \date{\today}

\begin{abstract}
We introduce and analyze a variational wave function for 
quasi two-dimensional $\kappa $-(ET)$_{2}$ organic salts 
containing strong local and nonlocal correlation effects. We find an 
unconventional superconducting ground state for intermediate
 charge carrier interaction, sandwiched between 
a conventional metal at weak coupling
and a spin liquid at larger coupling. 
Most remarkably, the excitation spectrum is dramatically renormalized and 
is found to be the driving force for the formation of the
unusual superconducting state.

\end{abstract}

\maketitle

\vskip2pc

The proximity to a Mott insulating phase affects the behavior of a correlated material in a fundamental way\cite{Anderson87,Kotliar88}.
Since most Mott insulators are also
magnetically ordered, it is often not clear whether unconventional behavior observed in nearby phases is due to the proximity to a Mott transition, to a magnetic phase transition, or a combination of both.
Ideal materials to sort this out are obviously spin liquids, 
systems where the Mott insulating phase is magnetically disordered. Recently, 
such behavior was found in the magnetically frustrated, insulating
organic charge transfer salt $\kappa $-(ET)$_{2}$Cu$_{2}$(CN)$_{3}$ 
by Shimizu \emph{et al.}\cite{Shimizu03}. No magnetic long-range order 
was found down to $T \approx 30 \, {\rm mK}$. While the low-$T$ 
susceptibility is suppressed, consistent with the opening of a spin gap, 
the observed power-law dependence, $1/T_1 \propto T^2$, of the NMR-spin 
liquid relaxation rate\cite{Shimizu03} supports gapless (possibly nodal) 
excitations of the spin liquid.
Under pressure, $\kappa $-(ET)$_{2}$Cu$_{2}$(CN)$_{3}$ becomes, 
like many similar $\kappa $-(ET)$_{2}$X 
systems\cite{Lang96,Ross98,Singleton02},
a superconductor\cite{Komatsu96}.

\begin{figure}[tbp]
\includegraphics[width=7.5cm]{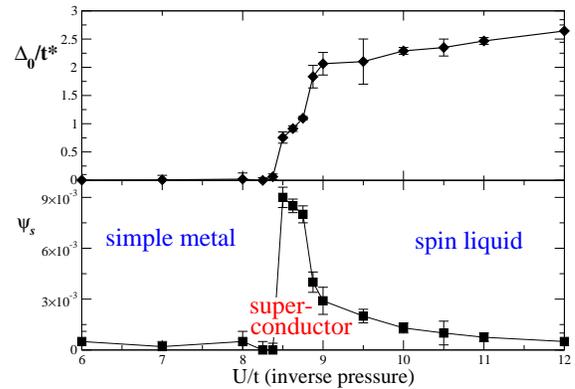}
\caption{Single particle gap, $\Delta_0$ in units of the renormalized
hopping $t^\ast$ (see text), and 
superconducting order parameter, $\psi_{\rm s}$, as a function of $U/t$. 
For $U/t \lesssim 8.5$ the system is a Fermi liquid metal.
Pairing and superconductivity build up in the 
intermediate coupling regime for $ U/t\sim 8 -10$ 
in a non-BCS fashion. 
While the pairing amplitude grows with $U/t$, superconductivity
is non-monotonic. For $U/t\gtrsim 10$,
$\psi_{\rm s}$ is negligible and the system becomes a spin liquid.
}
\label{figure01}
\end{figure}

In quasi two-dimensional $\kappa $-(ET)$_{2}$X 
salts\cite{Lang96,Ross98,Singleton02}
dimers of ET =
bis[ethylenedithio]-tetrathiafulvalene molecules are arranged in an 
anisotropic triangular lattice, with a charge state of one hole per dimer. 
The insulating state is, in most cases,
antiferromagnetically ordered and separated from a pressure induced 
superconducting phase by a first order transition\cite{Lefebvre}.       
In the superconducting state a number of experiments strongly support the 
existence of nodes of the pairing 
gap\cite{DeSoto95,Mayaffre95,Kanoda96,Nakazawa97,Belin98,Arai01,
Izawa02,comment01}. 
A gap with nodes was
determined in spin fluctuation
theories\cite{Schmalian98,Kino98,Kondo98,Vojta99,Kuroki02}; the location of the nodes of most of these calculations\cite{Schmalian98,Kino98,Kondo98,Vojta99}
was however in disagreement with experiments\cite{Arai01,Izawa02}. 
More importantly, the  absence of magnetic long range order in 
$\kappa $-(ET)$_{2}$Cu$_{2}$(CN)$_{3}$ and the rather strong first order 
transition to an antiferromagnet in other systems, seem to be at odds with 
key assumptions of the spin fluctuation approach. 

In this paper we introduce and analyze a variational 
resonating valence bond (RVB) wave function 
for systems close to a Mott transition, tuned by the interaction strength, 
i.e. for fixed carrier concentration but variable pressure.
Starting with a simple metal for low interaction strengths, we find a 
dramatic renormalization of the quasiparticle spectrum for intermediate 
strength of the interaction.
The system becomes a spin liquid of singlets with strong but short ranged 
spin correlations and a rapidly growing  gap which persists as the 
interaction strength is increased. 
We further find that
superconductivity is strong only in the transition regime between the 
simple metal and the spin liquid, see Fig.\ref{figure01}.
The nature of the renormalized spectrum and the nodal structure in the 
superconducting state are a strong function of the electronic dispersion.
We claim that our proposed wavefunction, with its ability to describe 
unconventional order parameters with nodes and effects of 
strong correlations, is a strong candidate for the description of 
$\kappa $-(ET)$_{2}$Cu$_{2}$(CN)$_{3}$ and related materials.

\begin{figure}[tbp]
\includegraphics[width=7.0cm]{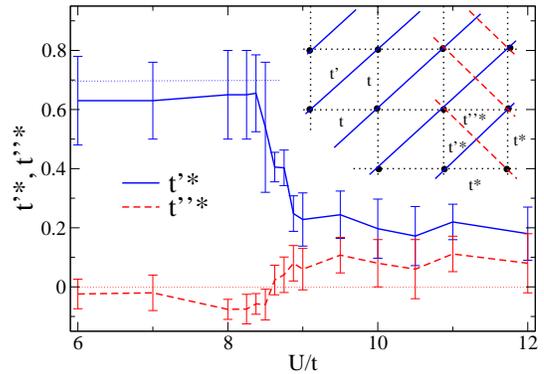}
\caption{The renormalized hopping elements along the 
two diagonals (see right part of the inset)
as a function of $U/t$. For $t^\prime=0.7 t$, the diagonal $t^{\prime\ast}$ is
strongly suppressed whereas $t^{\prime\prime\ast}$ is enhanced over their 
bare values (horizontal lines), suggesting a  reorganization of the quasiparticle dispersion 
for $U\sim 8.5t$. } 
\label{figure02} 
\end{figure}

We start from the single band Hubbard model 
\begin{equation} H=\sum_{ij;\sigma }t_{ij}c_{i\sigma }^{\dagger }
c_{j\sigma }+U\sum_{i}n_{i\uparrow }n_{i\downarrow }  
\label{Hubbard} \end{equation}
where $c_{i\sigma }^{\dagger }$ is the creation operator for a hole in 
the bonding state of a (ET)$_{2}$-dimer\cite{Ross98,Schmalian98,Chung01}. 
The hopping elements, $t_{ij}$, between dimers at sites $i$ and $j$ 
determine the bare band structure, 
$\varepsilon _{\mathbf{k}}=2t\left( \cos k_{x}+\cos
k_{y}\right) +2t^{\prime }\cos \left( k_{x}+k_{y}\right) $, 
as measured in magneto-oscillation experiments in the metallic 
regime\cite{Singleton02}, 
see inset of Fig.\ref{figure02}. 
$U $ is the Coulomb repulsion between holes on the same dimer. 
Typical parameters are $t\approx 0.05-0.1\mathrm{eV}$, and 
$U\simeq 5-10t$, whereas the ratio $t^{\prime }/t$ varies from 
$\simeq 0.6$ for $\ $X=Cu[N(CN)$_{2}$ ]Br to 
$t^{\prime }/t\simeq 1.0$ for the spin liquid compound with 
X=Cu$_{2}$(CN)$_{3}$.

In systems where the Mott transition occurs due to varying charge carrier 
concentration, important strong correlations are captured by the 
$t$-$J$ model. A great deal of insight into  this model has been gained 
by using a variational wave 
function\cite{Anderson87,Paramekanti01,Paramekanti04} 
$\left\vert \Psi _{\mathrm{RVB}}\right\rangle =e^{iS}P_{0}\left\vert 
\Phi _{\mathrm{BCS}}\right\rangle $, where $P_{0}$ projects out all 
doubly occupied states and $S$ generates the unitary transformation to the 
$t-J$ model (see 
Refs.\cite{Gros87,Paramekanti01,Paramekanti04,Anderson04,Watanabe04}). Here,
\begin{equation}
\left\vert \Phi _{\mathrm{BCS}}\right\rangle \propto 
\left( \sum_{\mathbf{k} }\varphi _{\mathbf{k}}
c_{\mathbf{k\uparrow }}^{\dagger }
c_{-\mathbf{k\downarrow }}^{\dagger }\right) ^{N/2}\left\vert 0
\right\rangle \end{equation} is a BCS wave function of $N$ particles with 
$ \varphi _{\mathbf{k}}=\Delta _{\mathbf{k}}/\left[ \xi _{\mathbf{k}}+
\sqrt{ \xi _{\mathbf{k}}^{2}+\Delta _{\mathbf{k}}^{2}}\right] $ and 
$\xi _{\mathbf{k}}=\varepsilon _{\mathbf{k}}-\mu _{\mathrm{v}}$. 
The gap $\Delta _{\mathbf{k}}$ as well as $\mu _{\mathrm{v}}$ are variational 
parameters\cite{Gros87,Paramekanti01,Paramekanti04,Anderson04,Watanabe04}.

In the organics the Mott transition occurs at half filling via changing 
pressure, i.e. changing the ratio $t/U$. Recently, a generalization of 
the $t-J$-model to pressure induced Mott transitions was suggested\cite{Baskaran03}. 
Here we use another approach and perform our calculation explicitly for 
finite $U/t$. This could be achieved using a Gutzwiller 
projected\cite{Gutzwiller65} pair wave function, 
$\left\vert \Psi _{\mathrm{GW}}\right\rangle =g^{D}\left\vert 
\Phi _{\mathrm{BCS}}\right\rangle $, with 
$ D=\sum_{i}n_{i\uparrow }n_{i\downarrow }$. Even though this wave function 
treats the interaction term in Eq.\ref{Hubbard} properly, it is known to 
poorly treat the kinetic energy and thus the physics of the superexchange 
coupling, $J=\frac{4t^{2}}{U}$. A promising way to improve this shortcoming  
was proposed in Ref.\cite {Kaplan82}. $g^{D}$ was replaced by 
$g^{D}h^{\Theta }$ with an additional variational parameter $h$. 
$\Theta $ is the total number of doubly occupied sites which have no 
empty neighbor connected by a hopping element. This causes an ``attraction" 
of doubly occupied and empty sites, that arise as intermediate steps for the 
superexchange process and dramatically improves the ground state 
energy\cite{Kaplan82}.

The propagation of quasi-particles in a magnetically frustrated system is 
likely affected by the strong, but local spin correlations and we expect 
that the renormalized spectrum, $\varepsilon _{\mathbf{k}}^{\ast }$, 
differs from the bare dispersion, $\varepsilon _{\mathbf{k}}$. 
This is a common observation of mean field or variational 
approaches\cite{Himeda00} to the $t-J$ model and was recently demonstrated 
in cluster dynamic mean field calculations of the present 
model\cite{Parcollet03}. Thus, we allow for a renormalization of the 
energy spectrum and use 
$\xi _{\mathbf{k}}=\varepsilon _{\mathbf{k}}^{\ast }-\mu _{\mathrm{v}}$ 
in the BCS-wave function. Since 
 $\left\vert \Phi _{\mathrm{BCS}}\right\rangle $ 
depends only on the ratio $\xi_{\mathbf{k}}/\Delta _{\mathbf{k}}$,
we cannot determine $\varepsilon _{\mathbf{k}}^{\ast }$ or the gap in
 absolute units, but, for example, in units of $t^\ast$,
the renormalization of $t$. $t^\ast$ itself cannot be determined.

Combining all these aspects, we propose the following wave function for 
organic charge transfer salts close to the Mott transition: 
\begin{equation} \left\vert \Psi \right\rangle =g^{D}h^{\Theta }
\left\vert \Phi _{\mathrm{BCS}}\right\rangle .  \label{wf} \end{equation}
The matrix element $E=\left\langle \Psi \left\vert H\right\vert 
\Psi \right\rangle /\left\langle \Psi |\Psi \right\rangle $ are evaluated 
using the Monte Carlo approach of Ref.\cite{Ceperley77,Paramekanti04} 
for a $12\times 12$ lattice. $E$ is  then minimized with respect to the 
variational parameters $h$, $g$, $\mu _{\mathrm{v}}$, $\Delta _{\mathbf{k}}$ 
and $\varepsilon _{\mathbf{k}}^{\ast }$ using a simulated annealing algorithm. 
The results shown below are mainly for $t^{\prime }/t=0.7$ and varying $U/t$, 
but we discuss other $t^{\prime }$-values as well.

{\em Correlated Superconductor}:
The ground state of the triangular lattice in the Heisenberg limit, $U/t\rightarrow \infty$ exhibits long range order\cite{Capriotti99}. 
From Ref.\cite{Morita02} we further know that for $t^{\prime }/t=0.7$ and
$U/t\lesssim 10-12$
the model is in a spin liquid insulating state without long range magnetic 
order.
We expect that such a spin liquid has a natural tendency towards 
Cooper pair formation and superconductivity\cite{Anderson87}. We calculate
the single particle gap $\Delta _{\mathbf{k}}$ and compare with the 
actual superconducting order parameter, $\psi _{s}$. 
Our central result is shown in Fig.\ref{figure01}a
depicting the $U$ dependence of the pairing amplitude of 
$\Delta _{\mathbf{k}}$. Out of a large class of different symmtries 
analyzed for $t^\prime=0.7t$, the optimal form of the gap is 
$\Delta_0 \left(\cos k_{x}-\cos k_{y} \right) $.
For small $U\lesssim 7t$, $\Delta _{0}$ is negligible; it starts building up for 
$U/t \sim 8.5$, rising sharply and attaining a large value of 
$\Delta_0/t^\ast \sim 3$.
Whether the pairing amplitude $\Delta_0$ is related to superconductivity is 
determined by the pairing correlation function, 
\begin{equation} 
F_{\mathbf{a,b}}\left( \mathbf{r-r}^{\prime }\right) =\left\langle 
\Psi \left\vert B_{\mathbf{r},\mathbf{a}}^{\dagger }B_{\mathbf{r}^{\prime },
\mathbf{b}}\right\vert \Psi \right\rangle , 
\end{equation} 
with $B_{\mathbf{r},\mathbf{a}}=\frac{1}{2}\left(
c_{\mathbf{r+a,\downarrow }
}c_{\mathbf{r\uparrow }}-c_{\mathbf{r+a,\uparrow }}c_{\mathbf{r,\downarrow }
}\right) $. 
In a superconductor,
$F_{\mathbf{a,b}}\rightarrow \pm \left\vert \psi _{s}\right\vert ^{2}$ 
for large $\left\vert \mathbf{r-r}^{\prime }\right\vert $. The sign depends on the direction of the nearest neighbor vectors $\mathbf{a}$ and $\mathbf{b}$. Our results for $\left\vert \psi _{s}\right\vert ^{2}$ as function of $U/t$ are shown in Fig.\ref{figure01}b. The order parameter becomes rapidly large around $U\simeq 8.5t$, just where $\Delta _{0}$ grows. 
For larger values of $U$, superconductivity becomes weak, despite the fact that $\Delta _{0}$ keeps growing. As shown in Ref.\cite{Millis91}, a wave function 
of the form in Eq.\ref{wf}, is unable to yield a true insulating state and 
it is not clear whether the state for $U\simeq 10t$ is a fragile 
superconductor along the lines of Ref.\cite{Laughlin}
or an insulator. 
In this strict sense our wave function describes 
a fragile conductor rather than a spin liquid.  We believe, however, 
that this limitation is of minor importance as far as the 
distinctive features of the superconductivity are concerned:
the large value of the gap, the very fragile superfluid stiffness
and their very different dependences on $U/t$.
A robust superconducting ground state only exists in the intermediate
transition regime.

\begin{figure}[tbp]
\includegraphics[width=7.0cm]{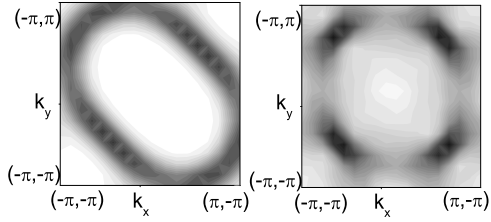}
\caption{Contour plot of $|\nabla n_{\bf k}|$ for $U=0$
(left) and for  $U/t=9$ (right). The strongly 
renormalized dispersion in the interacting case generates a four-fold
symmetric structure. $|\nabla n_{\bf k}|$ is largest along the diagonals.}
\label{figure03}
 \end{figure}

{\em Renormalization of the excitation spectrum}:
We find rather remarkably that the crucial factor for the emergence of the 
superconducting state is a qualitative change in the excitation spectrum.

As mentioned above,  we determine the optimal 
excitation spectrum, 
$\varepsilon _{\mathbf{k}}^{\ast }$ by minimizing the energy.
In addition to renormalizations of existing hopping elements, i.e. 
 $t\rightarrow t^{\ast }$ and  $t^{\prime }\rightarrow t^{\prime \ast }$, 
we include hopping elements 
not present in $\varepsilon _{\mathbf{k}}$. Allowing for all second and 
third neighbor hopping elements, 
one  relevant new  hopping  along the other diagonal 
$ t^{\prime \prime \ast }$ emerges (see inset of Fig.\ref{figure02}). 
Our results for  $t^{\prime \ast }$ 
and  $t^{\prime \prime \ast }$ are shown in Fig.\ref{figure02}. 
 A strong  renormalization 
$\varepsilon _{\mathbf{k}}\rightarrow \varepsilon _{\mathbf{k}}^{\ast }$ 
sets in coincident with the rapid increase of $\Delta_0$.
For $U\gtrsim 8t $, $t^{\prime \ast }$ is strongly reduced but the  additional hopping element, $t^{\prime \prime \ast }\sim t^{\prime \ast }$, develops.
As shown in Fig.\ref{figure03}, 
$ n_{\mathbf{k}}=\left\langle \Psi \left\vert c_{{\bf k}\sigma }^{\dagger } 
c_{{\bf k} \sigma} \right\vert \Psi \right\rangle $ 
becomes more symmetric than for the initial Hamiltonian and the effective 
dispersion corresponds to that of a weakly 
frustrated  system. At the crossover, $U/t\approx 8.5$, the non-local 
variational parameter $h$ rapidly decreases from $h \lesssim 1$ to
 small values, where the probability of doubly occupied sites being near an
 empty site is enhanced, thereby promoting superexchange physics.
Setting $h=1$ no gap or renormalized dispersion occurs  for $U/t<11$.
Our variational, $T=0$, calculations 
are completely consistent with the interesting effects recently found 
within a cluster dynamical mean field theory at finite $T$ by 
Parcollet {\em et al.}\cite{Parcollet03}, where for $U\approx 9t$
 new real space components of the self energy 
emerge  between precisely the lattice sites connected by 
$t^{\prime \prime \ast }$. 

\begin{figure}[tbp]
\includegraphics[width=7.0cm]{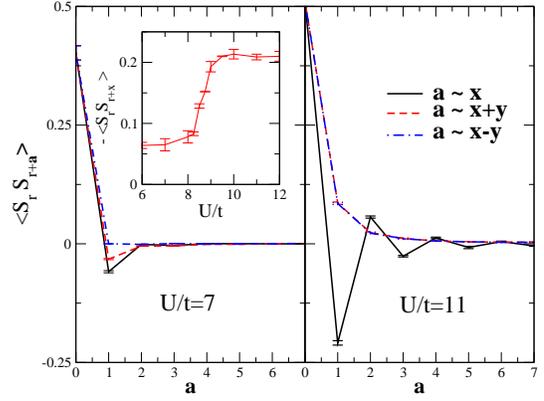}
\caption{Spatial dependence  of the spin-spin correlation function different $U$. For larger $U$ magnetic correlations along the two diagonals 
become indistinguishable even though the bare hopping elements
are very different.
The inset shows the $U/t$ dependence of the nearest 
neighbor spin correlation.} \label{figure04} \end{figure}

{\em Magnetic Correlations}:
The magnetic correlations of the spin liquid state at larger $U$ 
are strong but remain short ranged. This can be seen in 
Fig.\ref{figure04} where we show our results for the spin correlation function 
$\left\langle \Psi \left\vert s_{\mathbf{r}}^{\alpha }
s_{\mathbf{r}+\mathbf{a }}^{\alpha }\right\vert \Psi \right\rangle $ 
along different directions of $\mathbf{ a }$, for $U=7t$ and $U=11t$, 
respectively. For large $U$, there is, in agreement with the 
renormalized dispersion of Figs.\ref{figure02} and \ref{figure03},
almost no difference in the magnetic correlations along the two diagonals 
(essentially indistinguishable in Fig.\ref{figure04}), 
even though the bare hopping element in one direction is  zero. 
The state at large $U$  resembles the short range order of 
an \emph{unfrustrated} square lattice. The key difference  is of course 
that the square lattice for $U=11t$ is deep in the N\'{e}el ordered state, 
while the present model is in a spin liquid state\cite{Morita02}.
In the inset of Fig.\ref{figure04} we show the nearest neighbor spin 
correlation as function of $U/t$.

Finally we discuss the behavior for different $t'$-values, including the 
perfect triangular lattice at $t'=t$.
For $t'>t$, no new hopping element occurs, but
 $t'^*/t^*$ becomes much larger than its bare value.
 For $t'=1.5 t$ and  $U=12t$ we obtain $ t'^* \approx 7.5 t^*$. 
Again, an effective dispersion with weak frustration emerges;
now with a tendency to form weakly coupled chains.
 At the same time, the gap changes from $d$-wave to $\Delta_{\bf k}=\Delta_0 \left( \cos k_x+\cos k_y \right)- \Delta_0' \cos(k_x+k_y)$,
 with $\Delta_0 \sim \Delta_0'$. 
This corresponds roughly to a rotation of the nodes by $\pi/4$, 
as  found experimentally 
in organic superconductors\cite{Arai01,Izawa02}, however for a system 
where $t'$ is believed to be slightly smaller than $t$. 
For  $t' = t$, relevant for $\kappa $-(ET)$_{2}$Cu$_{2}$(CN)$_{3}$,
 a gap only occurs if we allow for a 
renormalization of the spectrum. 
Then, one of the three  bare hopping elements 
on the triangular lattice  is spontaneously reduced compared
 to the other two. 
A gap with equal size and opposite sign forms along those
 two nearest neighbor bonds with the larger effective hopping.
This demonstrate how crucial the renormalization of the
spectrum for pairing and supercondictivity is. It also shows 
that the physics of the triangular lattice at half filling but 
intermediate $U$ is  much richer than the behavior 
at weak coupling or in the strong coupling, Heisenberg limit. 

In summary, we have proposed a wave function for systems close to a pressure 
tuned Mott transition and  analyzed spin liquid formation and pairing 
in the ground state of  the Hubbard model on an anisotropic triangular lattice.
This geometry is relevant for quasi two-dimensional 
$\kappa $-(ET)$_{2}$X  organic superconductors. In the 
$\kappa $-(ET)$_{2}$Cu$_{2}$(CN)$_{3}$ member of this family, genuine spin 
liquid behavior was recently observed\cite {Shimizu03}. As a function of 
increasing correlation strength, corresponding to decreasing pressure, we find 
a rapid transitions between a weak coupling, 
simple metal regime and a 
strongly coupled spin liquid region with a number of interesting new 
properties, most remarkably a complete reorganization of the quasiparticle 
spectrum. The latter occurs in and close 
to the spin liquid regime and might be hard to 
observe via magnetooscillation 
experiments, but should affect optical or Raman experiments. 
 The large  gap of the spin liquid state has nodes, consistent with 
the observation of gapless modes in the low-$T$ spin lattice relaxation 
rate\cite{Shimizu03}.  Superconductivity emerges in the spin 
liquid - metal crossover  and is expected to become weak for larger pressure. 
The renormalized dispersion of the theory is either that of a square lattice 
or of weakly coupled chains, depending on the ratio $t^\prime/t$ of
 the hopping  elements.

We are grateful G. Kotliar and M. Randeria for discussions. 
This research was supported by the Ames Laboratory, operated for the U.S.
Department of Energy by Iowa State University under Contract No.
W-7405-Eng-82 (JL and JS). We acknowledge the use 
of computational facilities at the Ames Laboratory.

\end{document}